# Unexpectedly resisting protein adsorption on self-assembled monolayers terminated with two hydrophilic hydroxyl groups

Dangxin Mao, Yuan-Yan Wu[*], and Yusong Tu[*]

*College of Physics Science and Technology, Yangzhou University, Jiangsu 225009, China*

[*]Corresponding author: yywu@yzu.edu.cn; ystu@yzu.edu.cn

**Abstract**

The OH-terminated self-assembled monolayers, as protein-resistant surfaces, have significant potential in biocompatible implant devices, which can avoid or reduce adverse reactions caused by protein adhesion to biomaterial surfaces, such as thrombosis, immune response and inflammation. Here, molecular dynamics simulations were performed to evaluate the degree of protein adsorption on the self-assembled monolayer terminated with two hydrophilic OH groups ($(OH)_2$-SAM) at the packing densities ($\Sigma$) of 4.5 $nm^{-2}$ and 6.5 $nm^{-2}$, respectively. The results show that the $(OH)_2$-SAM itself can significantly improve the performance of its resistance to protein adsorption. This is attributed to the structure of the $(OH)_2$-SAM itself, i.e., the formation of a nearly perfect hexagonal-ice-like hydrogen bond structure in the OH matrix of the $(OH)_2$-SAM at $\Sigma = 4.5$ $nm^{-2}$, which sharply reduces the number of hydrogen bonds (i.e., 0.9) formed between the hydrophobic $(OH)_2$-SAM surface and protein. While for $\Sigma = 6.5$ $nm^{-2}$, the hydrophilic $(OH)_2$-SAM surface can provide more hydrogen bonding sites to form hydrogen bonds (i.e., 7.3) with protein. The number of hydrogen bonds formed between the $(OH)_2$-SAM and protein at $\Sigma = 6.5$ $nm^{-2}$ is ~8 times higher than that at $\Sigma = 4.5$ $nm^{-2}$, reflecting the excellent resistance to protein adsorption exhibited by the structure of $(OH)_2$-SAM itself at $\Sigma = 4.5$ $nm^{-2}$. Compared with traditionally physical barrier effect formed by a large number of hydrogen bonds between the $(OH)_2$-SAM and water above at $\Sigma = 6.5$ $nm^{-2}$, the structure of the $(OH)_2$-SAM itself at $\Sigma = 4.5$ $nm^{-2}$ proposed in this study significantly improves resistance to protein adsorption, which provides new insights into the mechanism of resistance to protein adsorption on the $(OH)_2$-SAM. These findings will be useful for designing protein-resistant materials with higher performance.

**Introduction**

The OH-terminated self-assembled monolayer (SAM), as one of the most promising resistance to protein adsorption surface, has great potential in medical implantable biomaterials among various protein-resistant surfaces.[1-11] It was well known that the adhesion of proteins to surfaces of medical materials could easily compromise the performance of diagnostic devices, cause the loss of the lifespan or efficacy for prostheses, biosensors, and contact lenses, and induce adverse reactions, including inflammation, infection, thrombosis, and biomaterial induced cancer.[12-14] Therefore, it is urgent to develop protein-resistant surfaces for medical materials. The OH-terminated SAM with good biocompatibility and biological inertness[15,16] has unique advantages in resisting protein adsorption, which has been extensively studied.[15-23] Experimental and theoretical studies showed that the OH-terminated SAM resistance to protein/peptide adsorption was the best among $CH_3$-, $CF_3$-, $NH_2$-, COOH-, and OH-terminated SAMs.[15-23] For the hydrophobic $CH_3$- and $CF_3$-terminated SAMs, the strong hydrophobic interactions could easily drive protein/peptide to directly adsorb on the SAMs due to no hydrogen bond (H-bond) formation between the hydrophobic SAM and water above, and thus the $CH_3$- and $CF_3$-terminated SAMs were not suitable as candidates for resistance to protein/peptide adsorption.[20-22] For the $NH_2$- and COOH-terminated SAMs with large charge distribution in the headgroups, although a large number of H-bonds were formed between the SAM and water above, the dominant electrostatic interactions between the SAM and protein/peptide could sufficiently compensate for the energy penalty to break H-bonds between the SAM and water above, leading to protein/peptide direct adsorption on the SAM without resistance to protein/peptide adsorption.[20,22,23] While for the OH-terminated SAM with weak charge distribution in the headgroups, a large number of H-bonds between the SAM and water above formed a physical barrier, and the weak protein/peptide-SAM interactions could not adequately overcome the physical barrier, making the OH-terminated SAM exhibit high resistance to protein/peptide adsorption.[20-23] As analyzed above, the OH-terminated SAM resistance to protein adsorption is traditionally attributed to the physical barrier effect formed by a large number of H-bonds between the SAM and water above, without considering the structure of the SAM itself. However, we note that the OH-terminated SAM itself can form a hexagonal-ice-like H-bond structure in the OH groups, exhibiting unexpected hydrophobicity in our previous study.[24] How does such a SAM structure affect resistance to protein adsorption? Thus, it is necessary to explore the effect of the structure of the OH-terminated SAM itself on resistance to protein adsorption to improve the performance of the SAM resistance to protein adsorption.

In this study, we will perform molecular dynamics (MD) simulations to evaluate the degree of protein adsorption on the $(OH)_2$-SAM at the packing densities of 4.5 $nm^{-2}$ and 6.5 $nm^{-2}$, respectively. Through our results, we expect to provide new insights into the mechanism of resistance to protein adsorption on the $(OH)_2$-SAM and the design of protein-resistant materials with higher performance.

**Model and method**

Model

The SAM consisted of the same number of five-carbon long alkyl chains terminated with two hydrophilic OH groups. The model atoms at the other end of the alkyl chains were bound to a (111) facet of a face-centered cubic (FCC) lattice. The packing densities ($\Sigma$) were 4.5 nm$^{-2}$ and 6.5 nm$^{-2}$, corresponding to the hydrophobic and hydrophilic structures in our previous study,[24] respectively. The crystal structure of protein was taken from the Protein Data Bank (PDB code: 6HNF). The protein solvated by the SPC/E[25] water molecules was initially placed at 6 Å above the (OH)$_2$-SAM. Sodium and chloride ions were added to neutralize the system with an ionic concentration of 0.15 M.

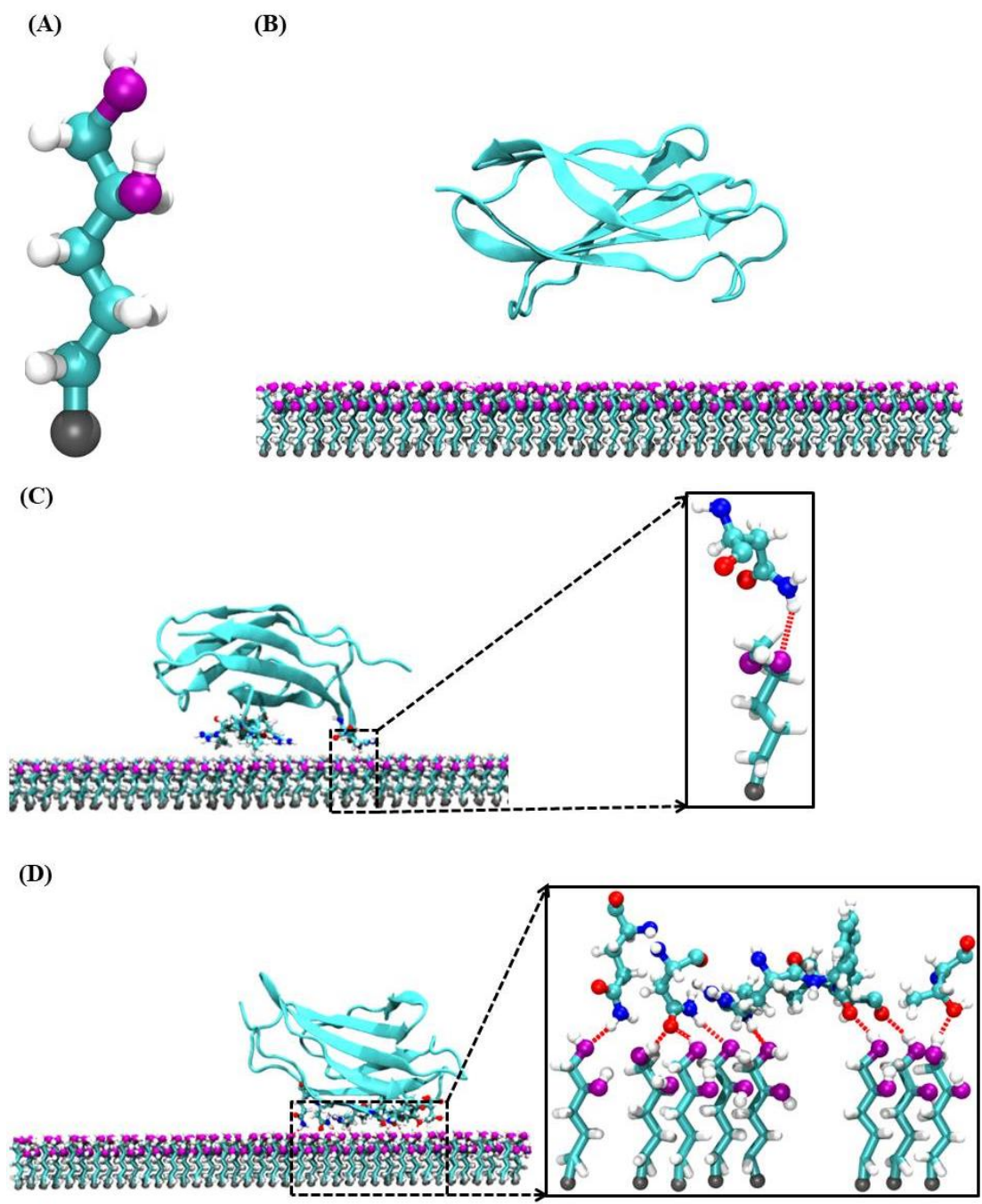

**Figure 1.** (A) Configuration diagram of a single alkyl chain. Color codes: hydroxyl groups, purple and white; main chain, cyan and white; and model atoms, gray. (B) The initial configuration of protein (NewCartoon mode) adsorption on the $(OH)_2$-SAM. The rest of the system is omitted for clarity. The final snapshots of the side view of protein adsorption on the $(OH)_2$-SAM at $\Sigma = 4.5$ nm$^{-2}$ (C), 6.5 nm$^{-2}$ (D). The H-bonds formed between the protein residues (ball and stick mode) and the alkyl chains are shown in red dashed lines in the insets.

Equilibrium MD simulations

MD simulations were performed using GROMACS 2019.6 package[26,27] with an OPLS-AA force field.[28] Each system was equilibrated in the canonical (NVT) ensemble using velocity rescaling thermostat at room temperature for 600 ns. The last 20 ns trajectories were used for analysis. Periodic boundary conditions were used in all directions. Particle-mesh Ewald (PME) algorithm[29,30] was used to treat long-range electrostatic interactions with a 1.0 nm cut-off, likewise, a van der Waals cut-off of 1.0 nm was used. All bonds involving hydrogen atoms were constrained using the LINCS algorithm.[31] Newton's equations of motion were integrated by a leap-frog algorithm with a 2 fs time step. The coordinate data were saved every 5 ps.

Steered MD simulations and umbrella sampling simulations

The structures from the equilibrium stage were used as the starting configurations for further steered MD simulations. The distance between the center of mass of protein and the model atoms of the $(OH)_2$-SAM was defined as a reaction coordinate, varying from ~2 nm to ~5 nm, where protein was pulled away from the $(OH)_2$-SAM along the z-axis (normal to the $(OH)_2$-SAM surface). Approximately 40 umbrella windows from steered MD trajectories were selected to further umbrella sampling with a window spacing of ~0.1 nm. Each umbrella window was equilibrated for 1 ns, followed by a 5 ns production simulation using a spring constant of 2000 kJ mol$^{-1}$ nm$^{-2}$ and a pull rate of 0.01 nm ps$^{-1}$. The potentials of mean force (PMFs) were calculated using the Weighted Histogram Analysis Method (WHAM).[32] The PMF in the aqueous phase was chosen as a reference point and defined to zero. The binding free energy was defined as the difference between the highest and lowest value of the PMF profile.

**Results and discussion**

To assess the convergence of the MD simulations, the root mean square deviations (RMSDs) of the protein backbone (relative to the crystal structure) at the two packing densities and in pure water were calculated, respectively (Figure S1, Supporting Information). The results show that the RMSD values of all three systems are stable around 0.18 nm throughout MD simulations, indicating that proteins are well equilibrated and do not have a large conformational deviation. The secondary structure changes of proteins were analysed using DSSP program.[33] As shown in

Figure S2 of Supporting Information, the main secondary structures of protein, i.e., 7 β-sheets, were well preserved during the simulation, indicating that the protein structure is stable and the native conformation is not disrupted.

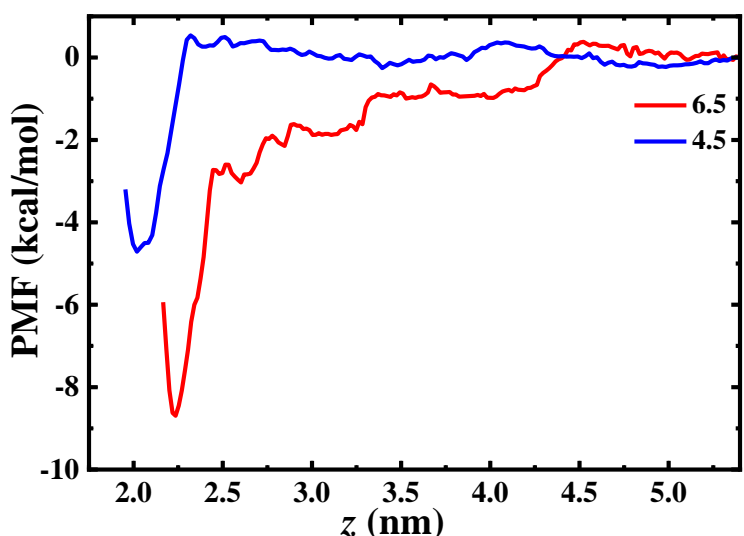

**Figure 2.** The PMF profile for protein along the reaction coordinate (z-axis) at $\Sigma$ = 4.5 nm$^{-2}$ (blue line) and 6.5 nm$^{-2}$ (red line), respectively.

The binding affinity of protein on the (OH)$_2$-SAM surface was analyzed by PMF, as shown in Figure 2. It can be clearly observed that the binding free energy at $\Sigma$ = 6.5 nm$^{-2}$ is more negative compared to that at $\Sigma$ = 4.5 nm$^{-2}$ with a difference of ~ -4 kcal mol$^{-1}$, indicating the stronger binding affinity of protein on the (OH)$_2$-SAM surface at $\Sigma$ = 6.5 nm$^{-2}$, while the binding affinity was significantly weaker at $\Sigma$ = 4.5 nm$^{-2}$. This is attributed to the effect of the structure of the (OH)$_2$-SAM itself, i.e., the formation of a nearly perfect hexagonal-ice-like H-bond structure (Figure 4) in the OH matrix of the (OH)$_2$-SAM at $\Sigma$ = 4.5 nm$^{-2}$ resulted in fewer H-bonds (~0.9, Table 1) between the (OH)$_2$-SAM and protein at $\Sigma$ = 4.5 nm$^{-2}$. Therefore, the binding affinity is weaker. For $\Sigma$ = 6.5 nm$^{-2}$, the hydrophilic surface formed more H-bonds (~7.3, Table 1) with protein, leading to the stronger binding affinity. This contrast indicates the strong resistance to protein adsorption due to the structure of the (OH)$_2$-SAM itself at $\Sigma$ = 4.5 nm$^{-2}$.

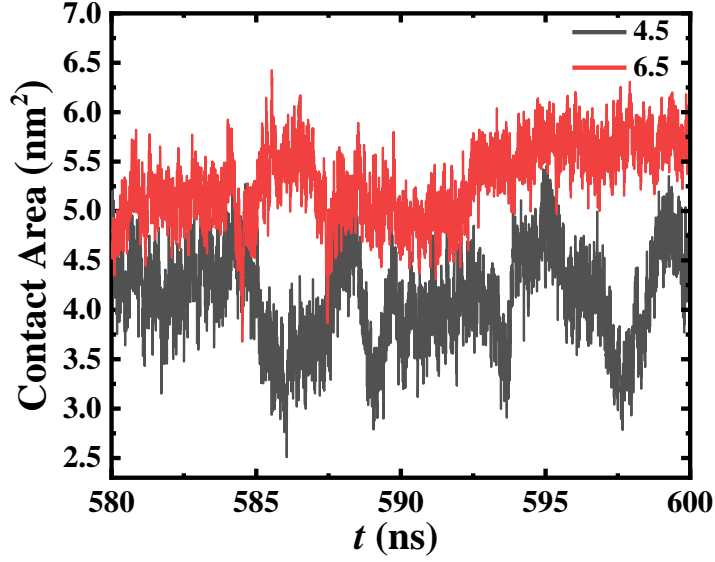

**Figure 3.** The contact area between protein and the $(OH)_2$-SAM surface at $\Sigma = 4.5$ nm$^{-2}$ (black line) and 6.5 nm$^{-2}$ (red line) for the last 20ns, respectively.

The contact area between protein and the $(OH)_2$-SAM surface can further assess the extent of protein adsorption on the $(OH)_2$-SAM surface, which is the area of the molecular surface buried in contact surface between two molecules. It was calculated as below[34-36]

$$\text{contact area} = \frac{1}{2}\left(\left(\text{SASA}_{protein} + \text{SASA}_{surface}\right) - \text{SASA}_{complex}\right) \quad (1)$$

where $\text{SASA}_{protein}$ and $\text{SASA}_{surface}$ are solvent accessible surface area of the isolated protein and $(OH)_2$-SAM, respectively, and $\text{SASA}_{complex}$ is that of protein and the $(OH)_2$-SAM complex. A probe of 1.4 Å radius is used to calculate the individual solvent accessible surface area of protein, the $(OH)_2$-SAM, and the complex.

It can be seen from Figure 3 that the contact area between protein and the $(OH)_2$-SAM surface at $\Sigma = 4.5$ nm$^{-2}$ fluctuates drastically, indicating the weak adsorption between protein and the $(OH)_2$-SAM surface. Conversely, the contact area at $\Sigma = 6.5$ nm$^{-2}$ is larger than that at $\Sigma = 4.5$ nm$^{-2}$ with smaller fluctuation, reflecting the stronger adsorption of protein on the $(OH)_2$-SAM surface. The different extent of protein adsorption at the two packing densities is attributed to the effect of the structure of $(OH)_2$-SAM itself at $\Sigma = 4.5$ nm$^{-2}$, leading to significant resistance to protein adsorption. Thus, the adsorption of protein on the $(OH)_2$-SAM surface is weak at $\Sigma = 4.5$ nm$^{-2}$. While the hydrophilic $(OH)_2$-SAM surface at $\Sigma = 6.5$ nm$^{-2}$ provides more H-bond sites to form H-bonds with protein, resulting in the stronger adsorption of protein on the $(OH)_2$-SAM surface and the larger contact area.

**Table 1.** H-bond number analysis at $\Sigma = 4.5$ nm$^{-2}$ and 6.5 nm$^{-2}$

| $\Sigma$ (nm$^{-2}$) | SAM-SAM | SAM-water | SAM-protein |
|---|---|---|---|
| 4.5 | 8.2 | 2.8 | 0.9 |
| 6.5 | 2.0 | 7.0 | 7.3 |

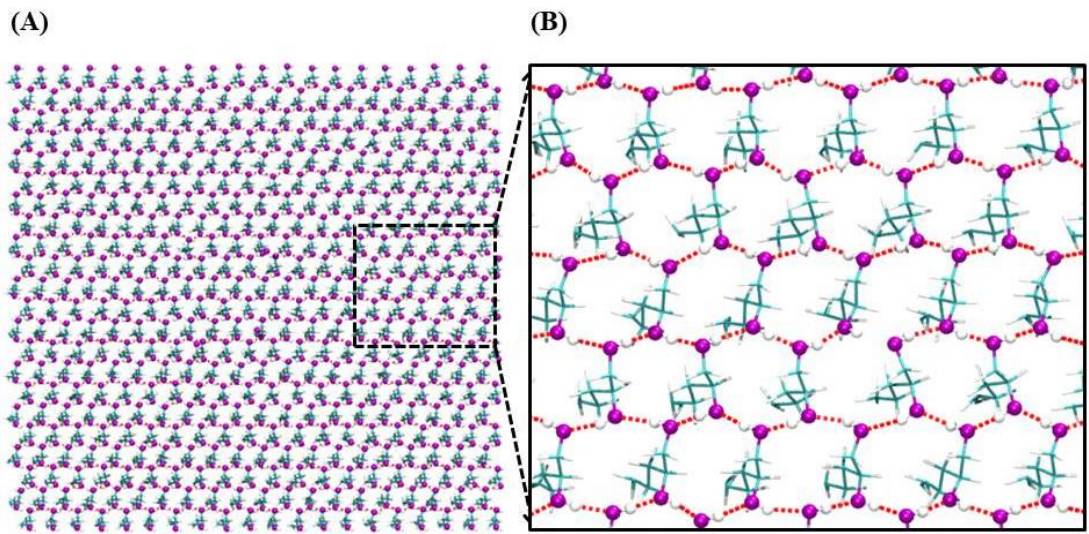

**Figure 4.** (A) The simulation snapshot of the top view of the (OH)$_2$-SAM at $\Sigma = 4.5$ nm$^{-2}$, (B) and the subfigure with the hexagonal-ice-like H-bond structure in the OH matrix of the (OH)$_2$-SAM. The atom representations and color settings are the same as the previous.

To further evaluate the strength of protein adsorption on the (OH)$_2$-SAM and the reasons for the difference in adsorption, we analyzed the average number of H-bonds among the OH groups of the (OH)$_2$-SAM (SAM-SAM, Table 1), between the (OH)$_2$-SAM and water above (SAM-water, Table 1), and between the (OH)$_2$-SAM and protein (SAM-protein, Table 1). As shown in Table 1, the structure of (OH)$_2$-SAM itself at $\Sigma = 4.5$ nm$^{-2}$ with the H-bond number of ~8.2 nm$^{-2}$ (SAM-SAM) leads to strong resistance to protein adsorption. It is shown that the number of H-bonds between protein and the (OH)$_2$-SAM is ~0.9, indicating that the (OH)$_2$-SAM resistance to protein adsorption is significantly excellent at $\Sigma = 4.5$ nm$^{-2}$. On the contrary, the hydrophilic (OH)$_2$-SAM surface with more H-bond sites and protein formed ~7.3 H-bonds at $\Sigma = 6.5$ nm$^{-2}$, leading to the stronger adsorption of protein on the (OH)$_2$-SAM. The number of H-bonds formed between the (OH)$_2$-SAM and protein at $\Sigma = 6.5$ nm$^{-2}$ is ~8 times higher than that at $\Sigma = 4.5$ nm$^{-2}$, reflecting unexpectedly resisting protein adsorption exhibited by the structure of the (OH)$_2$-SAM itself at $\Sigma = 4.5$ nm$^{-2}$. Meanwhile, we observe that the H-bond number between the (OH)$_2$-SAM and water above at $\Sigma = 6.5$ nm$^{-2}$ is ~2.5 times that at $\Sigma = 4.5$ nm$^{-2}$, indicating that a large number of H-bonds were formed between the (OH)$_2$-SAM and water above at $\Sigma = 6.5$ nm$^{-2}$. A physical barrier formed by a large number of H-bonds between the (OH)$_2$-SAM and water above is traditionally considered to be responsible for resistance to protein adsorption. However, our analysis shows that resistance to protein adsorption at $\Sigma$ = 4.5 nm$^{-2}$ is much stronger than that at $\Sigma$ = 6.5 nm$^{-2}$, which is also confirmed by the weaker electrostatic interactions between the (OH)$_2$-SAM and protein at $\Sigma$ = 4.5 nm$^{-2}$ in Figure S3 of Supporting Information. That is, resistance to protein adsorption exhibited by the structure of the (OH)$_2$-SAM itself at $\Sigma$ = 4.5 nm$^{-2}$ is much stronger than that exhibited by the physical barrier formed by a large number of H-bonds between the (OH)$_2$-SAM and water above at $\Sigma$ = 6.5 nm$^{-2}$.

**Conclusions**

In this study, the degree of protein adsorption on the (OH)$_2$-SAM surface was evaluated using MD simulations. The PMF showed that the binding affinity of protein on the (OH)$_2$-SAM was significantly weaker at $\Sigma$ = 4.5 nm$^{-2}$ with a difference of ~ -4 kcal mol$^{-1}$ compared to that at $\Sigma$ = 6.5 nm$^{-2}$, which was consistent with the smaller and more oscillating contact area of protein on the (OH)$_2$-SAM surface at $\Sigma$ = 4.5 nm$^{-2}$. We found that the structure of (OH)$_2$-SAM itself played a key role in resistance to protein adsorption, i.e., the (OH)$_2$-SAM itself formed a nearly perfect hexagonal-ice-like H-bond structure in the OH matrix of the (OH)$_2$-SAM at $\Sigma$ = 4.5 nm$^{-2}$, which sharply reduced the number of H-bonds between the (OH)$_2$-SAM and protein. In contrast, the hydrophilic (OH)$_2$-SAM surface formed more H-bonds with protein at $\Sigma$ = 6.5 nm$^{-2}$, resulting in the stronger protein adsorption on the (OH)$_2$-SAM. The number of H-bonds formed between the (OH)$_2$-SAM and protein at $\Sigma$ = 6.5 nm$^{-2}$ is ~8 times higher than that at $\Sigma$ = 4.5 nm$^{-2}$, reflecting unexpectedly resisting protein adsorption exhibited by the structure of the (OH)$_2$-SAM itself at $\Sigma$ = 4.5 nm$^{-2}$. Our analysis shows that resistance to protein adsorption exhibited by the structure of the (OH)$_2$-SAM itself at $\Sigma$ = 4.5 nm$^{-2}$ is much stronger than that exhibited by traditionally physical barrier effect formed by a large number of H-bonds between the (OH)$_2$-SAM and water above at $\Sigma$ = 6.5 nm$^{-2}$, which reflects that the structure of (OH)$_2$-SAM itself can significantly improve the performance of the (OH)$_2$-SAM resistance to protein adsorption. Our study will provide new insights into the mechanism of resistance to protein adsorption on the (OH)$_2$-SAM and the design of protein-resistant materials with higher performance.

**Acknowledgments**

This work was supported by the National Natural Science Foundation of China (Grants No. 12075201), the Science and Technology Planning Project of Jiangsu Province (BK20201428), the Postgraduate Research & Practice Innovation Program of Jiangsu Province (KYCX21_3193), and the Special Program for Applied Research on Supercomputation of the NSFC-Guangdong Joint Fund (the second phase).

**Conflicts of Interest**

The authors have no conflicts to disclose.